\documentclass[preprint]{revtex4}
\usepackage{graphicx}
\usepackage{dcolumn}
\usepackage{bm}
\begin{document}
\title{Classical tests of general relativity in the Newtonian limit
of Schwarzschild-de Sitter spacetime}
\author{ H.~Miraghaei$^{a}$\footnote{
Electronic address:~miraghaei@phymail.ut.ac.ir} and M.~Nouri-Zonoz $^{a,b,c}$ \footnote{
Electronic address:~nouri@theory.ipm.ac.ir, corresponding author}}
\address{$^{a}$ Department of Physics, University of Tehran, North Karegar Ave., Tehran 14395-547, Iran. \\
$^{b}$ School of Astronomy and Astrophysics, Institute for Research in
Fundamental Sciences (IPM), P. O. Box 19395-5531 Tehran, Iran.\\
$^{c}$Institute of Astronomy, Madingley road, Cambridge, CB3 OHA, UK.}
\begin{abstract}
Recently it has been shown that despite previous claims the
cosmological constant affects light bending. In the present
article we study light bending and the advance of Mercury's
perihelion in the context of the Newtonian limit of the
Schwarzschild-de Sitter spacetime employing the special
relativistic equivalence of mass and energy. In both cases,
up to a constant factor, we find the same results as in the
full general relativistic treatment  of the same phenomena. 
These approximate and intuitive arguments demonstrate clearly 
what effects should have been expected from the presence of $\Lambda$ 
in the general relativistic treatment of these phenomena.

\end{abstract}
\maketitle
\section{Introduction}
Recent cosmological observations point towards an accelerated
expansion of the Universe at the present epoch. In the
cosmological scale it is expected or otherwise speculated that the
cosmological constant is responsible, through a repulsive force,
for such an expansion. One might wonder why the same repulsive
force should not be at work in the solar system scale. In
particular, why $\Lambda$ should not produce such a repulsive
force around a local source of a spherical spacetime such as Sun
leading, in principle, to very small deviations from the well known
predictions of GR based on Schwarzschild spacetime calculations.
The basis of this expectation is the fact that, as for
the usual Einstein field equations, the modified equations (those in
the presence of the cosmological constant) should also be employed
in both solar as well as  cosmological scales. It seems that it 
was in response to such a
question that Rindler and Ishak have recently \cite{Rindler}
revisited the question of light bending in Schwarzschild-de Sitter
(SdS) spacetime and find out that despite previous claims
\cite{Islam} \cite{lake} $\Lambda$ affects light bending. Since
then there has been  a lot of discussion /controversy specially
over employing the same results in a consistent manner to the
cosmological lensing problem in the context of Freedman-Lemaitre
models \cite{Gibbons} - \cite{Park}. Specially it was shown by
Gibbons et al. why the absence of $\Lambda$ in the equation of null
geodesics in SdS spacetime does not imlpy that light bending is
independent of $\Lambda$. They find the origin of this absence in
the fact that in finding the null geodesic equation one projects
them onto a hypersurface of constant time. In this way the
projective properties of the null rays are independent of the
cosmological constant but other geometrical or conformal
properties of the rays such as their lengths and angles should be
identified through the full metric where $\Lambda$ appears
\cite{Rindler}. It is an obvious fact that if the effect of a
parameter in the Newtonian limit of a metric is present in a
phenomenon (such as light bending) then the same effect should
also be present in the full general relativistic treatment of
the same phenomenon. It is also well known that, using intuitive arguments 
based on the relativistic equivalence of mass and energy in the one 
hand and Newtonian gravity on the other hand, one could obtain basic features of 
inherently general relativistic effects such as gravitational redshift 
and light bending \cite{Sexl-Hakim}. In what follows  we will employ 
the same strategy to find the effects of $\Lambda$ on Mercury's perihelion 
precession and light bending in the context of the Newtonian limit of SdS spacetime.
\section{Newtonian limit of the Schwarzschild-de Sitter spacetime}
Schwarzschild- de Sitter (SdS) spacetime in its static
form (or in the static patch) is given by the
following line element
\begin{eqnarray}
ds^2= f(r) dt^2 - \frac{1}{f(r)} dr^2 - r^2 (d\theta^2+\sin^2\theta d\phi^2)\\
f(r)= (1-\frac{2GM}{r} - \frac{\Lambda c^2}{3} r^2).
\end{eqnarray}
It is an spherically symmetric spacetime with two parameters $M$
and $\Lambda$ corresponding to mass and the cosmological constant
respectively. It reduces to Schwarzschild (de Sitter) spacetime
for $\Lambda =0$ ($M=0$) and obviously is interpreted as the de
Sitter spacetime in which a spherically symmetric point mass is
embedded. Since it is spherically symmetric one could read out the
Newtonian potential from the $g_{tt}$ component of the metric as 
(bringing in the fundamental constants $c$ and $G$);
\begin{eqnarray}\label{potential}
\Phi \approx -\frac{GM}{r} -\frac{1}{6}\Lambda r^2 c^2.
\end{eqnarray}
Since SdS is not asymptotically flat this is valid as long as $\Phi \ll 1 $
\footnote {This non-asymptotic flatness property of SdS spacetime is the main issue raised in \cite{Rindler}
to account for the contribution of $\Lambda$ to the light bending. Indeed this fact will 
link our result on light bending to that found in \cite{Rindler}. We refer the reader to the discussion after Eq. (27).}.
Indeed (\ref{potential}) is a solution to the modified Poisson's equation 
$\nabla^2\Phi = 4\pi G\rho_M - \Lambda c^2$ \cite{Harv}. Obviously the above
modified potential leads to the modified force (per unit mass);
\begin{eqnarray}\label{force}
F \approx  -\frac{GM}{r^2} + \frac{1}{3}\Lambda r c^2,
\end{eqnarray}
in other words the extra "harmonic oscillator type" potential in
(\ref{potential}) leads to the repulsive force $\frac{1}{3}\Lambda
r c^2$ (for $\Lambda > 0$). It is quite amazing to know that Newton in his discussion
on gravitational force in Principia, exploring other central force laws, had
come to the conclusion that the other central force for which one could consider the
whole mass of a spherically symmetric distribution at its center is a linear central force \cite{Newton}.
It seems that Pierre Simon Laplace was the first to  consider the superposition of the two
forces explicitly in the form $F =Ar^2 +\frac {B}{r^2}$ \cite{north}.
Looking at the Newtonian limit of SdS, the extra linear term contemplated by
Newton, has been noticed as a classical analogue to the introduction of the cosmological
constant term in general relativity \footnote{For a historical
background on this term refer to \cite{Lahav}.}.
It should also be noted that the compromise between $\Lambda$'s cosmological role and its role in an exact
solution for a compact object such as Sun is not trivial. We will get back to this point later in the article.
Our starting point to study classical
tests of general relativity in the context of  the Newtonian limit of SdS spacetime is the above modified force.
\section{Effect of $\Lambda$ on Mercury's perihelion precession}
Since $\Lambda$ appears explicitly in the equation of timelike
geodesics in SdS spacetime, unlike light bending there has been no
ambiguity over its effect on Mercury's perihelion precession.
Indeed using timelike geodesics in SdS one can easily show that
$\Lambda$ has a definite effect on  Mercury's perihelion precession in the form of an additional
advancement by an amount \cite{Rindlerbk}
\begin{eqnarray}\label{peri1}
\delta \Psi_{\Lambda} \approx  \frac{\pi \Lambda c^2 a^3
(1-e^2)^3}{GM_\odot},
\end{eqnarray}
in which $M_\odot$ is the Sun's mass (i.e the mass parameter in Schwarzschild metric) and $a$ and $e$ are the orbit's semi-major axis and eccentricity respectively \footnote{In \cite{mash} the same effect is considered to linear order in $\Lambda$ and the result is found to have a different dependence on the eccentricity of the form $(1-e^2)^{1/2}$. This does not make any difference in our case since we are going to consider only approximately circular orbits i.e $e\approx0$.}.
Here we aim to show how one can obtain the same result, up to a constant factor, by employing the above modified Newtonian gravity combined with special relativistic mass-energy equivalence, namely $E = Mc^2$, applied to both the planet and the Sun's {\it modified} gravitational field.
First of all we note that from special relativity for a moving mass such as a planet we have
\begin{eqnarray}
m={\frac {m_{{0}}}{\sqrt {1-{\frac {{v}^{2}}{{c}^{2}}}}}},
\end{eqnarray}
in which $m_0$ is the planet's rest mass. Secondly the gravity field around sun has energy and correspondingly a mass which for the usual gravitational potential in (\ref{potential}) is given by \cite{Sexl-Hakim}
\begin{eqnarray}
M_{g}=\frac{1}{2}\frac{{M_\odot}^2 G}{r c^2}.
\end{eqnarray}
Calculating the same kind of mass for the linear part of the gravitational potential (\ref{potential}) corresponding to the cosmological constant $\Lambda$, we note that from Poisson's equation
\begin{eqnarray}
\nabla^2 {\Phi_\Lambda} = - 4\, \pi \, G \rho_\Lambda,
\end{eqnarray}
one can assign a uniform energy density $\rho_\Lambda = \frac{\Lambda c^2}{4\pi G}$  and consequently, through mass-energy equivalence, a mass
\begin{eqnarray}
M_{\Lambda}={\int_0^r} \rho_\Lambda d V^{\prime} = {\int_0^r} \rho_\Lambda 4\pi {r^{\prime}}^2 d r^{\prime} =\frac{\Lambda c^2}{3G}r^3,
\end{eqnarray}
which will affect a planet in a circular orbit of radius $r$ around the Sun. Applying the above relations to the total energy of a planet at radius $r$ around sun we end up with
\begin{eqnarray}\label{6}
E = (m-m_0) c^{2}-{\frac { G m M(r)}{r}}-\frac{1}{6}\Lambda m r^2c^2,
 \end{eqnarray}
 in which
 \begin{eqnarray}
M(r) = M_\odot+ M_\Lambda + M_g = M_\odot+\frac{1}{3}\,{\frac {\Lambda\,{c}^{2}{r}^{3}}{G}}+\frac{1}{2}\,{\frac
{{M_\odot}^{2 }G}{r{c}^{2}}},
 \end{eqnarray}
is the Sun's modified mass at a distance $r$. It should be noted that $\Lambda$ appears twice in Eq. (10). Once through its contribution in the modified mass and then through its own potential energy arising from the $\Lambda$-dependent gravitational potential in Eq. (3). Substituting for $m$ and $M(r)$ in (\ref{6}) and expanding $m$ in terms of $\frac{v}{c}$ and using the fact that for an approximately circular orbit 
$\frac{mv^2}{r}\approx{\frac {{GmM}}{r^2}}$, we find up to the second order in $\frac{v}{c}$,
\begin{eqnarray}
E \approx \frac{1}{2}m_0{v}^{2}-\frac { G m_0 M_\odot}{r}-{\frac
{5{G}^{2}m_0{M_\odot}^{2}}{8{r}^{2 }{c}^{2}}}-\frac{1}{2}{
m_0 r}^{2}\Lambda\,{c}^{2}+\frac{1}{6} m_0 M_\odot G\Lambda\,r,
\end{eqnarray}
in which the first two terms in the right hand side are the familiar Newtonian contribution to the planet's energy. The third term is due to the mass attributed to the gravitational energy $M_g$ in Eq. (7) and through this crude reasoning, responsible for the precession of Mercury's perihelion \cite{Sexl-Hakim} giving both the order of magnitude and the correct sign of the effect as compared to the general relativistic value \footnote{There is a point need to be mentioned here. Comparing the term found here with that in \cite{Sexl-Hakim} (Eq. (4.20) in Hakim's book) we note that there is a difference in the numerical factor and we obtain $\frac{5}{8}$ instead of $\frac{3}{4}$.
This difference arises from the wrong expression used for the kinetic energy of the planet in \cite{Sexl-Hakim}, resulting in the wrong second order term  ($\frac{v^2}{c^2}$ ) in the expansion of the  kinetic energy.}. The last two terms correspond to $\Lambda$'s contribution. Ignoring the last term in comparison to the fourth term (since their ratio  $\frac{GMm/r}{mc^2}\ll 1$) it is noted that the extra advancement of planet's perihelion due to the cosmological constant is given by the ratio $\frac{\phi_{\Lambda}}{\phi_{N}}$ in which $\phi_{\Lambda}$ and ${\phi_{N}}$ are the $\Lambda$-dependent  and Newtonian gravitational potential energies (i.e the fourth and the  second terms in the right hand side of Eq. (12)) respectively, so that;
 \begin{eqnarray}
{\frac {\delta \Psi_\Lambda}{2\pi }}={\frac {\phi_{{\Lambda}}}{\phi_{{N}}}}
 =\frac{1}{2}\frac {\Lambda\,{r}^{3}{c}^ {2}}{G{M_\odot}},
 \end{eqnarray}
 or
 \begin{eqnarray}
\delta \Psi_\Lambda \approx {\frac {\pi \,\Lambda\,{r}^{3}{c}^{2}}{G{M_\odot}}}.
 \end{eqnarray}
Comparing this result with (\ref{peri1}) for an approximately circular orbit (i.e $e\approx0$) it is surprising that we have found the exact general relativistic result through the above approximate approach. It has already pointed out in the literature that the above result could only put an upper limit on the value of $\Lambda$ of order $10^{-42} cm^{-2}$, since for values larger than this the effect of the cosmological constant on Mercury's perhelion would be observable \cite{Rindlerbk}. But we know from the cosmological observations that this upper limit has already been well respected and the value of the cosmological constant is $14$ orders of magnitude smaller. In other words the precession induced by $\Lambda$ is at least $10^{14}$ times smaller than that due to the gravitational energy given by the ratio $\frac{\phi_{GE}}{\phi_{N}}$, in which $\phi_{GE}$ is the third term in 
the right hand side of Eq. (12).
\section{Effect of $\Lambda$ on Light bending}
The general classical equation of trajectory for a particle of mass $m$ and angular momentum  $l$ under the action of  a central force $F(r)$ is given by \cite{Gold};
\begin{eqnarray}\label{1}
{\frac {d^{2}}{d{\phi}^{2}}}u \left( \phi \right) +u=-{\frac {mF\left( \frac{1}{u} \right)
 }{{l}^{2}{u}^{2}}},
 \end{eqnarray}
in which $u=\frac{1}{r}$.
To study bending of light in a gravitational force field of the form (\ref{force}), we follow Laplace and assign a mass $m_\gamma$ to photon so that the above equation takes the following form
 \begin{eqnarray}\label{2}
{\frac {d^{2}}{d{\phi}^{2}}}u \left( \phi \right)
+u={\alpha}^{-1}-{ \frac {\delta}{{u}^{3}}},
 \end{eqnarray}
in which $\delta={\frac {{\it \Lambda c^2 m_{\gamma}^2 }}{3{l}^{2}}}$ and
$\alpha={\frac {{l}^{2}}{{\it GMm_{\gamma}^2}}}$ \footnote{Note that we have applied the force per unit mass (\ref{force}) on the mass $m_\gamma$.}.
To solve the above equation we use a perturbation method based on the fact that the last term in the above equation is much smaller than the other terms due to the smallness of $\delta$. Therefore, to the first order in $\delta$, we assume a solution of the form
 \begin{eqnarray}\label{3}
u=u_{{0}}+\delta\,u_{{1}},
 \end{eqnarray}
in which $u_{{0}}={\frac {1+\epsilon\,\cos \phi }{\alpha}}$ is the zeroth order solution corresponding to the orbit of a particle under the influence of the usual inverse square law with eccentricity $\epsilon= \sqrt{1+\frac {2E\alpha}{mMG}}$. Now substituting (\ref{3}) in (\ref{2}), we find the following equation for $u_1$
 \begin{eqnarray}
\frac{d^2}{d\phi^2} u_1 (\phi)
+u_1 = -\frac {\alpha^3}{(1+\epsilon\,\cos
\phi)^3},
 \end{eqnarray}
whose special solution, for $\epsilon \gg 1$ (i.e $\epsilon \approx \sqrt{\frac {2E\alpha}{mMG}}$) is given by;
 \begin{eqnarray}\label{03}
u_{{1}}\approx \frac{\alpha^3}{2\epsilon^2}\left({\frac {2\, \cos ^{2} \phi
 -1}{1+\epsilon \cos \phi }}\right),
 \end{eqnarray}
and consequently the general solution for $u$ is as follows;
 \begin{eqnarray}\label{4}
u \approx {\frac {1+\epsilon\,\cos  \phi 
}{\alpha}}+\frac{1}{2}\,{\frac {\delta{\alpha}^{3} \, \left( 2\, 
\cos^{2}\phi -1\right) }{\epsilon^2 (1+{\epsilon}\cos  \phi)}}.
 \end{eqnarray}
Now the bending angle is given by,
 \begin{eqnarray}\label{5}
\gamma=2\,\phi_{{\infty }}- \pi,
 \end{eqnarray}
in which $\phi_{\infty }$ is the particle's (here Photon's) azimuthal 
angle (measured from the axis joining the deflector to the point of closeset approach) at infinity. It is found by setting $r=\infty$ or equivalently $u=0$ in (\ref{4});
\begin{eqnarray}\label{7}
\phi_{\infty } \approx {\it Arccos} (-\frac{1}{\epsilon}\pm \sqrt
{\frac {\delta}{2}} \frac{\alpha^2}{\epsilon^2}),
\end{eqnarray}
to the lowest order in $\delta$ and in $\frac{1}{\epsilon}$. Again expanding the above function we have;
\begin{eqnarray}
\phi_{\infty} \approx \frac{\pi}{2} + \frac{1}{\epsilon}- \sqrt
{\frac{\delta}{2}}\frac {\alpha^2}{\epsilon^2},
\end{eqnarray}
where we have chosen the plus sign in (\ref{7}) to avoid the singularity in (\ref{4}) at $\phi = \frac{\pi}{2} +  \frac{1}{\epsilon}$.
Substituting the above result in (\ref{5}) we end up with;
\begin{eqnarray}
\gamma=\frac{2}{\epsilon} - \sqrt
{2\delta}\frac{\alpha^2}{\epsilon^2}.
\end{eqnarray}
It is noted that the first term is the usual bending angle due to the inverse square law which is half the general relativistic value as expected. The second term which is induced through the linear force term in (\ref{force}) bears the repulsive nature of $\Lambda$ in being negative and obviously reduces bending. This repulsive nature of $\Lambda$ could also be seen through the angular dependence of the extra ($\Lambda$-dependent) term (\ref{03}) added to the light path in the absence of the cosmological constant. It is equal to zero at $\phi=\frac{\pi}{4}$ (where $r=r_0$) and maximum at $\phi=0$, so it is negative for $0 < \phi < \frac{\pi}{4}$ and positive for $\frac{\pi}{4} < \phi < \frac{\pi}{2} +  \frac{1}{\epsilon}$. This leads, for the same angular positions, to larger (smaller) radii for the perturbed path compared to the unperturbed path for $\frac{\pi}{4} < \phi <  \frac{\pi}{2} +  \frac{1}{\epsilon}$ ($0 < \phi < \frac{\pi}{4}$). Consequently the tangents to the unperturbed path will have steeper slopes compared to the perturbed path and thereby leading to a larger bending (Fig. 1). It is also interesting to note that the perturbed path is dimpled around the closest approach due to this extra term. This corresponds, in the general relativistic treatment \cite{Rindler}, to the spatial projection of the light path on the Flamm praboloid near the Schwarzschild region of SdS (where $\frac{m}{r}$ dominates over $\Lambda r^2$)
as comapred to the projection on a sphere with radius $\sqrt{\frac{3}{\Lambda}}$ in its de Sitter region (where  $\Lambda r^2$ dominates over $\frac{m}{r}$).
To write the bending angle in terms of more common parameters we note that photon's angular momentum is given by $l=m_{\gamma} c R$ in which $R$ is the distance of the closest approach. Also taking $E \approx \frac{1}{2} m_{{\gamma}}{c}^{2}$  as photons energy by neglecting its potential energy and substituting for $l$ in $\alpha$  we find
 \begin{eqnarray}
\epsilon \approx {\frac {{c}^{2} R}{{\it GM}}},
 \end{eqnarray}
and consequently the contribution of $\Lambda$ to the bending angle is given by
 \begin{eqnarray}
\gamma_{\Lambda} \approx -\sqrt{\frac{2\Lambda}{3}} R.
 \end{eqnarray}
Being dependent on the square root of $\Lambda$, and the fact that from cosmological observatons $\Lambda \approx 10^{-56}  cm^{-2}$, the above value is more pronounced compared to the general relativistic value obtained by Rindler and Ishak,
\begin{eqnarray}
\gamma_{\Lambda} \approx -\Lambda\frac{c^2R^3}{6 M G},
\end{eqnarray}
depending on $\Lambda$ itself \cite{Rindler}. This difference on the dependence of bending angle on the cosmological constant originates from the fact that, working in the Newtonian realm, we have allowed $r\longrightarrow \infty$ in (\ref{4}) which is obviously not allowed in full SdS spacetime. In fact it was this non-asymptotic flatness of SdS spacetime which led Rindler and Ishak to a local definition of bending angle for different angular positions ($\Phi$) of the observers \footnote{To avoid any ambiguity between the two angles defined here and in \cite{Rindler}, we denote their azimuthal angle by the capital letter $\Phi$.}. To show the effect of this important fact more clearly, we find the bending angle for
the finite value of $r$ corresponding to $\phi=\frac{\pi}{2}$, which is geometrically equivalent to the intrinsically characterized $r$ value at $\Phi=0$ in \cite{Rindler}. We do this by computing the slopes of the tangents to the perturbed and unperturbed paths at $\phi=\frac{\pi}{2}$. These are given by 
\begin{eqnarray}
\beta_0 \approx \tan\beta_0 = \left(\frac{1}{r_0}\frac{dr_0}{d\phi}\mid_{\phi=\frac{\pi}{2}}\right)^{-1}=\frac{1}{\epsilon},
\end{eqnarray}
and
\begin{eqnarray}
\beta \approx \tan\beta = \left(\frac{1}{r}\frac{dr}{d\phi}\mid_{\phi=\frac{\pi}{2}}\right)^{-1}\approx \frac{1}{\epsilon}
(1 - \frac{3}{2} \frac{\delta \alpha^4}{\epsilon^2}),
\end{eqnarray}
for the unperturbed and perturbed paths respectively (Fig. 1). Therefore the contribution of the cosmological constant to the bending angle in this case is given by
\begin{eqnarray}
\gamma_\Lambda \approx 2(\beta - \beta_0) \approx -3\frac{\delta \alpha^4}{\epsilon^2}= -\Lambda \frac{R^3 c^2}{M G},
\end{eqnarray}
which compared to the general relativistic value Eq. (27), not only has the right sign but also, up to a constant factor of $\frac{1}{6}$, is the same $\Lambda$-dependent result. Again as in the case of the Mercury's perihelion, due to the very small value of the cosmological constant, it is expected that the effect of $\Lambda$ on light bending to be negligible. Indeed for a light ray grazing Sun's limb the light bending induced by the cosmological constant is $10^{28}$! times smaller than that by the Sun's mass itself \cite{Rindler}.
\begin{figure}
\begin{center}
\includegraphics[angle=0,scale=0.7]{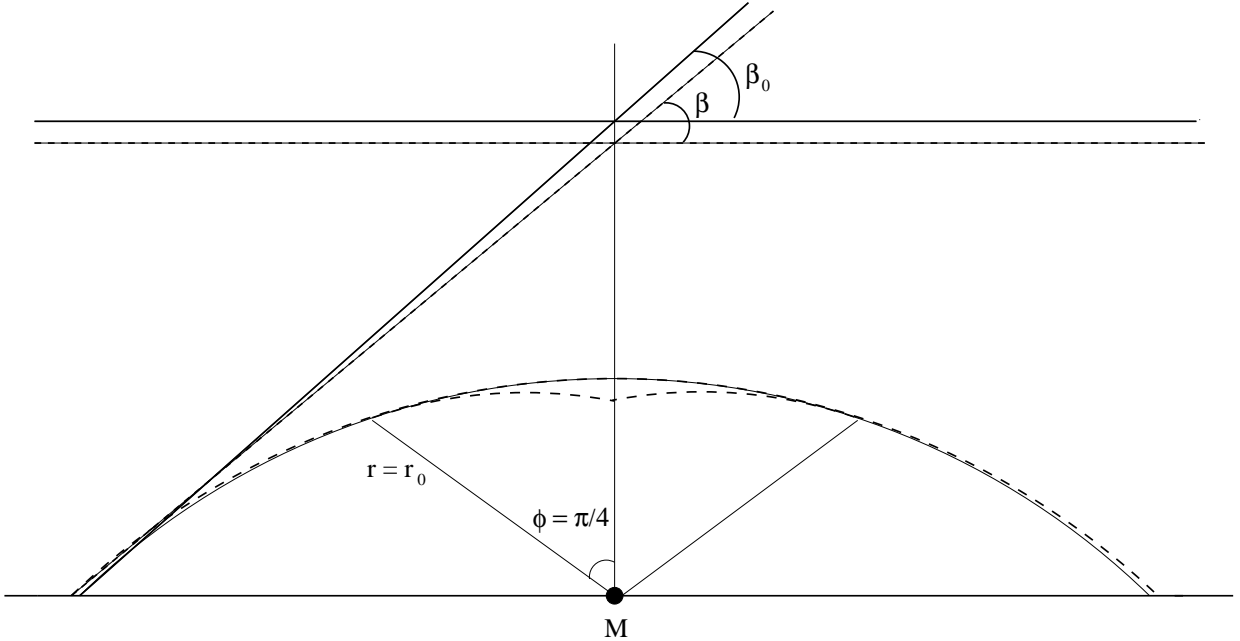}
\caption{Light paths in the presence (dashed line; $r(\phi)$) and in the absence (solid line; $r_0(\phi)$ ) of $\Lambda$ for $-\frac{\pi}{2}< \phi < \frac{\pi}{2}$. Note that the difference between the paths is enormously exaggerated due to the small value of $\Lambda$.  It is also intersting to note that the perturbed path (dashed line) has the same shape (in being dimpled around the closest approach) as the spatially projected light path ${\cal L}^2$ in Fig. 1 of Ref. \cite{Rindler}.}
\end{center}
\end{figure}
\section{discussion}
The above results show clearly that one should have expected the effect of $\Lambda$ on light bending in a general relativistic treatment of the same phenomenon based on the Schwarzschild-de Sitter spacetime as the metric around a spherical deflector such as Sun.\\
As pointed out in the introduction one could think of the above calculation as the calculation of light bending in Newton's modified gravity in which a linear force is added to the inverse square law. But then in the purely Newtonian case one should be careful with the form of this extra term if one way or another it is going to be related to the Newtonian limit of an appropriate general relativistic spacetime. For example
in \cite{Lahav}, as an straightforward conjecture, the full force (per unit mass) due to a point mass $M$ according to Newton's modified gravity is given by $F=-\frac{GM}{r^2} + CMr$ in which $C$ is a constant. In other words the extra (linear) force is taken to be originated from the  body's mass $M$ and proportional to it. It is obvious that now  $CM$ should be related to the cosmological constant if one is looking for a central repulsive force in a cosmology based on this modified Newtonian gravity. To do so the authors have used the Friedmann equation in the presence of $\Lambda$ in the weak field limit
\begin{eqnarray}
\frac{\ddot{a}}{a} = -\frac{4\pi G \rho}{3} + \frac{\Lambda}{3},
\end{eqnarray}
applied to a sphere of radius $r$ and mass $M$ leading to
\begin{eqnarray}
{\ddot{r}}= -\frac{GM}{r^2} + \frac{\Lambda}{3}r,
\end{eqnarray}
so that $\frac{\Lambda}{3}= CM$. In this way $\Lambda$, in the cosmological scale, is related to the mass of the universe in the same spirit as in the Einstein static Universe \footnote{Recall that in Einstein static Universe $\Lambda = \frac{4\pi G \rho}{c^2}\propto \frac{GM}{c^2 R^3}$ in which $M$ and $R$ are the mass and radius of the Universe respectively.} and of course in line with Mach's ideas on the origin of inertia. But it should be noted that this form for the linear force is not compatible with the Newtonian limit of SdS spacetime (as an exact  solution of Einstein modified field equations for a compact spherical object of mass $M$) since, unlike the gravitational force in the Newtonian limit of SdS (Eq. (4)), here by setting the deflector mass $M$ equal to zero the total gravitational force vanishes. This originates in the simple fact that setting $M=0$, SdS spacetime reduces to that of de Sitter and not Minkowski  i.e $\Lambda$ in SdS spacetime is a new parameter independent of the mass of the central object.
\section *{Acknowledgments}
M. N-Z thanks D. Lynden-Bell for useful discussions and comments. The authors would like to thank University of Tehran for supporting this project under the grants provided by the research council.

\end{document}